\documentstyle[12pt]{article}
\newcommand{\p}{\underline}
\newcommand{\be}{\begin{equation}}
\newcommand{\ee}{\end{equation}}

\newcommand{\th}{\Theta}
\newcommand{\im}{\imath}

\newcommand{\fr}{\frac}

\newcommand{\ep}{\epsilon}
\newcommand{\vph}{\varphi}

\newcommand{\pr}{\partial}
\newcommand{\kp}{\kappa}

\newcommand{\al}{\alpha}
\begin{document}
\title{THE TIME SURFACE TERM IN QUANTUM GRAVITY}
\author{V.N. Pervushin, V.V. Papoyan, G.A. Gogilidze,\\
        A.M. Khvedelidze, Yu.G. Palii, V.I. Smirichinskii\\
        Joint Institute for Nuclear Research,\\
        Dubna, 141980, Moscow region, Russia\\
        email: pervush@thsun1.jinr.dubna.su}
\maketitle
\medskip
PACS number(s):04.60.-m, 04.20.Cv, 98.80.Hw (Quantum Gravity)
\medskip

\begin{abstract}
The role of the time surface term in the ADM Hamiltonian formulation of
general relativity is investigated. We show that the variable contained
in the time surface term (the scale factor) plays the role of a
time-like variable. The conjugated variable represents the energy density
in the reduced phase space, where
the Schr\"odinger like equation for a wave function is derived.
The contribution from the surface term to the phase of the wave function
allows us to define {\bf the phase time of
the quantum Universe} so that it coincides with
{\bf the proper time as an invariant interval} for the classical dust filled
Universe. The quantum scenario of the evolution of the Universe filled in by
the Weinberg-Salam  fields is considered.
The wave function  of the early  Universe as the functional
from  the Higgs  fields and scale factor realizes  the unitary irreducible
representation of the SO(4,1) group. The elementary particle masses
are determined by the angles of the scale--scalar field mixing.
\end{abstract}

\vspace*{1.5cm}
\noindent
\section{Statement of the problem}
The Dirac-ADM canonical approach \cite{Dir58,ADM}  to GR was the essential
improvement
on the way to quantization of the Einstein-Hilbert theory of gravity
developed by Wheeler, DeWitt and others \cite{Wheel,DeWitt,Mis,Ryan}.
This conventional scheme of the canonical quantization is based on the
(3+1) Dirac-ADM foliation
\cite{Dir58,ADM,kuchar,Zel}  of the four dimensional manifold
$(x^\mu)$ along the some  time-like vector (associated with
the rest frame of an observer):
$$
ds^2=N^2dt^2-{}^{(3)}g_{ik}\breve d x^i \breve d x^k \;\;;\;\;
(\breve d x^i=dx^i + N^idt)\,
$$
(that means the restriction  of  the group of general coordinate
transformations by the kinemetric ones
\cite{Zel}: $t\to t'(t)\;\;\;;\;\;\;x_i\to {x'}_i(t,x_1,x_2,x_3)$)
and on the ADM  action ($W_{ADM}$) which differs from the initial
Einstein-Hilbert action \cite{Hil15}
\be \label{action}
W_{GR}=\int d^4x\sqrt{-g}\left [-\frac{{}^{(4)}R(g)}{2\kappa^2} +
         {\cal L}_{matter} \right ]\;\;;\;\;(\kappa^2=8\pi G),
\end{equation}
by the surface terms:
$$
W_{GR}=W_{ADM}+W_S+W_T,
$$
where
\begin{equation} \label{t}
W_T=-\int dtd^3x {\breve \partial}_0 \left [\frac{{\breve \partial}_0
\sqrt{{}^{(3)}g}}{N\kappa^2} \right ]\;\;;\;\;\left [{\breve \partial}_0f=
\dot f -\partial_k (N^kf)\right].
\end{equation}

\begin{equation} \label{s}
W_S=-\int dtd^3x \partial_k[\sqrt{{}^{(3)}g}({}^{(3)}g^{ik}
\partial_l N)]\frac{1}{\kappa^2}\;.
\end{equation}
For the derivation of local classical equations these surface terms are
not essential, however, they play an important role in the determination
of the global quantities of such a total energy \cite{Regge}.
It is obvious that the wave function also has global nature and
depends on these surface terms.
{\bf The statement of the problem consists in the canonical quantization of
the Einstein-Hilbert action~(\ref{action}) by taking into account the
surface terms (\ref{t}),~(\ref{s})}. We continue the attempts to solve this
problem in the papers \cite{Perv,Khved}.

\section{A new version of the Hamiltonian Formulation}

{\bf To solve this problem we should

I) consider  the space-scale variable $a=[{}^{(3)}g]^{\frac{1}{6}}$
in the time surface term~(\ref{t})}  as one of  dynamical variables :
\be \label{metric}
ds^2=a^2[N_c^2dt^2-\omega_{i\p k}\omega_{j\p k}{\breve d}x^i\breve d x^j],
\;\;\;;\;\;\;aN_c=N\;\;\;;\;\;\; {\rm det}\omega=1,
\end{equation}
(we use the triad form $\omega_{i\p j}$ for the rest  dynamic
variables~\cite{Smir})

{\bf II) apply the Ostrogradsky method} ~\cite{Ostr,gkp}
to the theory with second order derivative of the scale factor with respect
to the time coordinate. As the result, the Hilbert action (\ref{action})
in terms
of the canonical conjugate variables $ a,\pi_a, \Phi,\pi_\Phi$
(where $\Phi$ denotes the set of matter fields
including graviton $\omega$ and photon $A$) has the form
$$
W_{GR}=\int^T_0 dt \int_V d^3 x \left [-\pi_{(a)}\stackrel{\odot}{a}+
\frac{1}{2}{\breve\partial}_0(\pi_{(a)}a)+\sum_{\Phi =\omega ,A} \pi_{(\Phi)}
\stackrel{\odot}{\Phi}
-N_c{\cal H}_{EC}\right] + W_S,
$$
where
$$
\stackrel{\odot}{a}=\breve\partial{}_0a+\frac{2}{3}a\partial_kN^k,\;\;\;\;\;
\stackrel{\odot}{A}_k=\partial_0A_k-\partial_kA_0-N^lF_{lk}
$$
$$
\stackrel{\odot}{\omega}_{lk}=(\dot\omega {}_{l\p s}\omega {}_{k\p s}
+\omega {}_{l\p s}\dot\omega {}_{k\p s}
-\nabla_lN_k -\nabla_kN_l+
\frac{2}{3}\omega_{l\p s}\omega_{k\p s}\partial_jN^j),
$$
is the kinemetric invariant time derivative (we use here covariant derivative
in the metric $\omega_{i\p k}\omega_{j\p k}$,
including the Laplace operator
$\Delta f=\nabla_k\partial^kf$), and ${\cal H}_{EC}$ is the Einstein energy
density
\be \label{eenergy}
{\cal H}{}_{EC}=-\frac{\kappa^2}{12}\pi^2_{(a)}+ \pi^2_{(\omega)}\left(
\frac{2\kappa^2}{a^2}\right)+\left(\frac{a^2}{2\kappa^2}\right)\bar R+
{\cal H}_{(A)};\;\;\;\;
\ee
with the photon energy
$$
{\cal H}{}_{(A)}= \frac{1}{2}\pi_{(A)}^k{\pi}_{k (A)}+\frac{1}{4}F_{ij}F^{ij}
\;\;
$$
and three dimensional curvature
$$
\bar R=a^2\;\;{}^{(3)}R(a^2\omega^2)={}^{(3)}R(\omega^2) +
8a^{-\frac{1}{2}}\Delta a^{\frac{1}{2}},
$$

{\bf III) perform the  canonical transformation}
$(\pi_{(a)},a) \ =>\;(\Pi,\eta)$
which removes the time surface term $\frac{1}{2}\partial_0(\pi_{(a)}a)$
\begin{equation}   \label{ct}
\pi_{(a)}\stackrel{\odot}{a}-\frac{1}{2}\breve\partial{}_0(\pi_{(a)}a)=
\Pi
(\dot\eta-N^k\partial_k\eta)\;.
\end{equation}
One can represent this transformation as
\begin{equation} \label{cantr}
\pi_{(a)}=2\sqrt{\frac{3\Pi}{\kappa^2\Gamma}} C(\eta);\;\;\;\;
a=\sqrt{\frac{\kappa^2\Gamma \Pi}{3}} S(\eta),
\end{equation}
where  $ C(\eta), S(\eta) $ and $\Gamma $ are some particular
solution of the following equations
\begin{equation} \label{cs}
 C(\eta)\frac{d}{d\eta} S(\eta)- S(\eta)\frac{d}{d\eta} C(\eta)=1;\;\;\;\;\;
\pr_0({\rm ln\Gamma})-N^k\partial_k({\rm
ln}\Gamma)+\frac{1}{3}\partial_kN^k=0\;.
\end{equation}
Finally, the Hilbert action reads in terms of the new variables as
\begin{equation} \label{ham}
W_{GR}=\int dtd^3x\left [\sum_{\Phi =\omega ,A,{\rm ln}\Gamma }  \pi_{(\Phi )}
{\dot \Phi }-\Pi {\dot \eta }
-N_c{\cal H}{}_{EC}-N^k{\cal P}_k \right]+{\bar W} {}_S\;,
\end{equation}
with space surface term
$\bar W{}_S=W_S-2\int^T_0dtd^3x\partial_l(N^k\pi_{(h)}{}^l_k)$
and  constraints
\begin{equation} \label{c1}
{\cal H}_{EC}=
-\frac{\Pi}{\Gamma}(C^2-S^2\frac{6}{\Gamma^2}\bar R)
+\pi^2_{(\omega)}\frac{6}{S^2\Pi\Gamma}+{\cal H}_{(A)}=0,
\end{equation}

\begin{equation} \label{c2}
{\cal P}_k=\pi_{(\Gamma)}\partial_k{\rm ln}\Gamma+\partial_k\pi_{(\Gamma)}+
   \Pi  \partial_k\eta +2\nabla_l\pi^l_{(\omega) k}+\pi_{(A)}^lF_{lk}=0.
\end{equation}

\section{Interpretation of new variables}

To  treat the new variables $\Pi $ and $\eta $, we consider the flat-space
limit \cite{Khved}:
$\pi_{(\omega)}=\bar R=N^k=0$;\\$C(\eta)=1;\;S(\eta)=\eta,$
where the Hilbert action~(\ref{ham}) has the form
$$
W_{GR}=\int dt d^3x\left\{\pi^k_{(A)}(\dot A_k-\partial_kA_0)-
\Pi\dot \eta
-N_c({\cal H}_{(A)}-\frac{\Pi}{\Gamma})\right\}.
$$
After the reduction on the constraint shell
$ {\cal H}_{EC} = 0$
in the gauge $N_c=1$, we get the  expression
for the conventional action of
electrodynamics
$$
W^{Red}=\int dt d^3x \left\{\pi^k_{(A)}(\dot A {}_k-\partial_kA_0)-
{\cal H}_{(A)}\right\},
$$
Note that because
\be
\label{em} 
\dot\eta=\frac{1}{\Gamma}.
\ee
{\bf in this limit  the variable  $\eta$ can be treated  as the time  and
              the quantity $\Pi /\Gamma $
in the reduced phase space is  a "reduced energy" like the
the quantity $ \sqrt{p^2+m^2} $ is  the spectral energy for a
relativistic particle.}

\section{The wave function of the Universe}

Suppose that  the constraint ${\cal H}_{EC}=0$ has the set of
solutions
 $\Pi={\cal H}^{red}_{(\alpha)}\;\;\;,\alpha=1...m$.
 For each solution one can write down the corresponding  reduced Hilbert
action
\be \label{wred}
W^{red}_{GR(\alpha)}=\int dt \int d^3x \left(\sum_{\Phi=\omega ,
A,{\rm ln}\Gamma} \pi_{(\Phi)}\dot\Phi
-{\cal H}^{red}_{(\alpha)}\dot\eta\right) +\bar W_S
\ee
The quantization of the action leads to the  Schr\"odinger type evolution
equation
\be \label{wdw}  
\frac{1}{i}\frac{\delta}{\delta\eta}\Psi_{\alpha}=\hat{\cal
H}{}^{red}_{(\alpha)}
\Psi_{\alpha} .
\ee
Let us consider the small time limit    $(\eta)\sim 0 $
which corresponds to the small Universe
 $(a\sim S(\eta)\to0)$.
 According to equation~(\ref{cs}), in this region $S(\eta)\sim\eta$,
$C(\eta)\sim 1 $, and the graviton term ${\pi }_{(\omega)}^2$
dominates in the energy density (\ref{c1}). The corresponding solutions of
the constraints ${\cal H}_{EC}=0$
 \be \label{pi} 
 \Pi={\cal H}^{red}_\pm = \pm\sqrt{6\pi_{(\omega)}^2}\frac{1}{\eta}.
\ee
 and  the reduced action  reads
 \be \label{mis} 
 W_{GR\pm} =\int^T_0dt\int_Vd^3x\left(\sum \pi_{(\omega)}\dot\omega
 \mp\sqrt{6\pi_{(\omega)}^2}\partial_0{\rm ln}\eta\right).
 \ee
 One can verify that in the supposition of
homogeneousity of the space
\begin{equation}  \label{homo}
 ds^2=a^2[N^2_c(t)dt^2-A^2(r)dx^2];\;\;\;
\;\;A(r)=\left(1+\frac{kr^2}{4r^2_0}\right)^{-1}
\end{equation}
\be \label{fried} 
a(t)N_c(t)dt=dT_{Fried};\;\;\;\;k=0,\pm1,
\ee
it follows  from (\ref{mis})  the action for the
Misner Universe \cite{Mis} (in details see~\cite{Ryan}):
 \be \label{1mis} 
 W_{GR\pm} = V_{(3)} \left(\pi_{(\omega)}(\omega(T)-\omega(0))\mp
 \sqrt{6\pi_{(\omega)}^2}{\rm ln}\frac{\eta(T)}{\eta(0)}\right),
 \ee
and the spectral decomposition for the wave function
\be  \label{psq} 
\Psi = \int d\pi_{(\omega)} \left( A^+_{\pi_{(\omega)}}e^{iW_{GR+}} +
A^+_{\pi_{(\omega)}}e^{iW_{GR-}} \right),
\ee
where the role of the time is played by the logarithm of $\eta$,
and $A^\pm$ are the operators of creation and annihilation of the Universe.
Note that in the  case of the homogeneous space the
 particular solutions of eq.~(\ref{cs}) read
\begin{equation} \label{cos}
 C(\eta)=1,{\rm cos{\eta}},{\rm cosh{\eta}}; \quad
 S(\eta)= \eta,{\rm sin{\eta}},{\rm sinh{\eta}},
\quad \Gamma =  r_0
\end{equation}
respectively for \(k = 0, 1, -1\).

The evolution of the Universe filled in by dust and radiation  has been
considered in \cite{Khved}.
In this case the wave function of the Universe has the
form $\Psi_\pm=\exp{\{\imath
W_\pm^{red}(a)}\}$  with the reduced action
$$
W^{red}_\pm=\pm V_{(3)} \int_{0}^{T}da \left[\p{\pi}_{(a)}-
  \fr{1}{2}\frac{d}{da}(\p{\pi}_{(a)}a)\right],
$$
where $V_{(3)}=\int d^3xA^3(r)$ and
$
\p{\pi}_{(a)}=2\sqrt{\frac{3}{\kp^2}}\left[{\cal H}_{(M)}-
   \frac{3ka^2}{r_0^2\kp^2}\right]^{1/2}
$
is the solution of the constraint (\ref{c1}) ${\cal H}_{EC}=0$. The phase
of this function coincides (up to the energy factor) with the Friedmann
time (\ref{fried}) for the dust case (${\cal H}_M=a\epsilon_{dust}$)
\be \label{d} 
W^{red}_{dust}=\frac{V_{(3)}\epsilon_{dust}}{2}T_{Fried}(a);\;\;\;\;\;\;\;\;
T_{Fried}(a)=\int_{a(0)}^{a(T)}da\frac{6a}{\kp^2 \p{\pi}_{(a)}}
\ee
and with the conformal time $N_cdt=\eta (t) r_0$ for the radiation
(${\cal H}_M= \epsilon_{rad} $)
\be \label{r} 
W^{red}_{rad}=V_{(3)}\epsilon_{rad}\eta (a)r_0;\;\;\;\;\;\;\;\;
\eta (a)r_0=\int_{a(0)}^{a(T)}da\frac{6}{\kp^2 \p{\pi}_{(a)}}.
\ee
It is worth to note that the this clear correspondence
between the quantum and classical physics ( the "phase time" and the
"interval time") arises due to the
maintenance of the  time surface term $\frac{1}{2}
\pr (\p{\pi}_{(a)}a)$ in the Hilbert action.

{\bf Thus, the time surface term helps us to establish the
correspondence between  the time as the phase of the ADM wave function of
the Universe and the classical proper time as an invariant interval.}

\section{Quantum scenario
of the Weinberg-Salam Universe.}

Let us consider the quantum scenario of the evolution of the Universe filled
in by the Weinberg-Salam  fields and described by the action
\begin{equation}     \label{ws}   
W=\int d^3xdt\sqrt{-g}\left\{-\fr{{}^{(4)}R}{2\kp^2}
+\fr{{}^{(4)}R \Phi^* \Phi }{6}+ \pr_{\mu}\Phi^* \pr^{\mu} \Phi
     -\gamma(\bar \Psi_{L}\Phi)\Psi_R+\cdots\right\}
\end{equation}
where $\Phi=\left(\begin{array}{c}\Phi_1\\
    \Phi_2
   \end{array}\right)$
is the doublet of the complex scalar fields; $\Psi_L$ and $\Psi_R$ are the left
and right fermions. We keep only the term of the scalar-fermion interaction
generating masses of the fermions \\
$m(\bar \Psi_{L2}\Psi_R+\bar \Psi_R\Psi_{L2})$
to show the evolution of the mass parameters with respect to the scale
$a$.
If we  extract the scale factor \( a \) not only from the metric
(\ref{metric})
$g_{\mu\nu}=a^2\tilde g_{\mu\nu}$, but also from all other
matter fields
 $\Phi=a\vph,\;\;\; \Psi=a^{-3/2}\Psi_c$, we can
guarantee the classical limit of the massive fermion fields as the
Friedmann dust of the Universe.
The action in terms of the physical fields $\vph,\Psi_c,\tilde g
(\sqrt{-\tilde g}=N_c)$ has the form
\begin{eqnarray}
W&=&\int d^4x\{N_c\left[-\frac{{}^{(4)}\tilde R}{2\kp^2}\left(a^2-
\frac{\kp^2}{3}\vph^*\vph\right)-\frac{3}{\kp^2}\pr_\nu a\pr^\nu a
+\pr_\nu \vph^*\pr^\nu\vph-\gamma(\bar
\Psi_{Lc}\vph^)\Psi_{Rc}+\cdots\right]\nonumber\\
&&+\pr_\nu\left(N_c\left[\frac{3a^2}{\kp^2}-\vph^*\vph
\right]\frac{\pr^\nu a}{a}\right)\}\nonumber
\end{eqnarray}
Note that the configuration   $a^2 =\frac{\kp^2}{3}\vph^*\vph $ represents
the singular  point: in the vicinity  of this point the sign
before the four- dimensional curvature  is changing.
So, let us consider only the field configuration  such that
$[a^2-\frac{\kp^2}{3}\vph^*\vph]=\rho^2 > 0$.
For this case one can introduce new variables
$$
 a=\rho \cosh(\xi);\;\;
 \vph_i=\sqrt{\frac{3}{\kp^2}}\rho \sinh(\xi) n_i;\;\;
n_1=\cos(\th)\exp{\{\im \chi_1\}};\;\;n_2=\sin(\th)\exp{\{\im \chi_2\}},
$$
where $\xi,\th,\chi_1,\chi_2$ are the angles of the scale-scalar mixing.

For the homogeneous space~(\ref{homo}) with $V_{(3)}=1$ we get the action
\begin{eqnarray}
 W&=&\int dt\{-\frac{D}{2N_c}\left(\frac{6}{\kp^2}\right)+
 N_c\left[\frac{k}{2r_0^2}\rho^2\left(\frac{6}{\kp^2}\right)
-\gamma\sqrt{\frac{3}{\kp^2}}\rho\sinh(\xi)(\bar
\Psi_{Lci}n_i)\Psi_{Rc}+\cdots\right]
\nonumber\\
&&+\pr_0\left[
 \frac{\rho^2}{2N_c}\left(\frac{\dot \rho}{\rho}+\dot \xi \tanh \xi\right)
\right]\frac{6}{\kp^2}\},\nonumber
\end{eqnarray}
where $D=-\dot \rho^2+\rho^2 \dot \xi^2+(\rho\sinh\xi)^2\left(\dot \th^2+
  \sin^2(\th)\dot \chi^2_1+\cos^2(\th)\dot \chi^2_2\right)$
is the $SO(4,1)$-invariant differential form.
By the Ostrogradsky method this action can be rewritten in terms of momenta
$$
W=\int dt\{\sum_{\al=\xi,\th,\chi_1,\chi_2}\pi_{(\al)}\dot \al-
\pi_{(\rho)}\dot \rho-
 N_c{\cal H}_{EC}+\frac{1}{2}\pr_0\left(\pi_{(\rho)}\rho\right)+
\frac{1}{2}\pr_0\left(\pi_{(\xi)}\tanh(\xi)\right)\}
$$
$$
{\cal H}_{EC}=-\frac{1}{2}\frac{\kp^2}{6}\pi^2_{(\rho)}-\frac{k\rho^2}{r^2_0}
\frac{6}{\kp^2}
+\frac{{\cal K}^2}{2\rho^2}\frac{\kp^2}{6}
+\gamma\sqrt{\frac{3}{\kp^2}}\rho\sinh(\xi)(\bar \Psi_{Lci}n_i)\Psi_{Rc}+\cdots
$$
where ${\cal K}^2$ is the Kazimir operator of the $SO(4,1)$ group
$$
{\cal K}^2=P^2_{(\xi)}+\frac{1}{\sinh^2(\xi)}\left(P^2_{(\th)}+
  \frac{P^2_{\chi_1}}{\sin^2(\th)}+\frac{P^2_{\chi_2}}{\cos^2(\th)}\right).
$$
We see that the variable $\rho$ plays the same role as the scale $a$ in the
theory without scalar fields, and it is the time like variable.
Note that our transition to new variables
is similar to the Bekenstein transformation~\cite{Beken}.
Repeating the canonical transformation (\ref{cantr})
 $\pi_{(\rho)}\dot \rho-
\frac{1}{2}\pr_0(\pi_{(\rho)}\rho)=\Pi\dot \eta$, where $\Gamma=r_0$, we get
the expression for the action
\begin{eqnarray}
W&=&\int dt\{\sum_{\al}P_{(\al)}\dot \al-\Pi\dot \eta-
   N_c\left[
-\frac{\Pi }{r_0}+\frac{{\cal K}^2}{4\Pi r_0S^2(\eta)}+
\gamma\sqrt{\frac{3}{\kp^2}}\rho\sinh(\xi)(\bar \Psi_{Lci}n_i)\Psi_{Rc}
+\cdots\right]\nonumber\\
&&+\frac{1}{2}\pr_0\left(P_{(\xi)}\tanh (\xi)\right)\}\nonumber
\end{eqnarray}
This action describes the following ADM-scenario of the evolution of the
Universe. In the small time limit $(\eta\sim 0)$ the Kazimir operator term
dominates and the reduced system on the constraint ${\cal H}_{EC}=0$ has
the form
$$
W^{red}_{(\pm)}=\int_{T_s(0)}^{T_s(T)}dT_s\left[\sum_{\al}
P_\al\frac{d\al}{dT_s}\mp{\cal
K}(P_\al)+\frac{1}{2}\frac{d}{dT_S}(P_\xi\tanh(\xi))
\right];\;\;\;\;dT_s=\frac{d\eta}{2S(\eta)}.
$$
The wave function of this system can be decomposed over the eigenfunctions
of the Kazimir operator with the eigenvalues ${\cal K}_
\epsilon$
$$
\Psi(T_s=T_s(T)-T_s(0)|\al)=\sum\limits_{\ep}
\left[A_\ep^{(+)}e^{+\im {\cal K}_\ep T_s}\Psi_\ep(\al_T|\al_0)
+A_\ep^{(-)}e^{-\im {\cal K}_\ep T_s}\Psi_\ep^*(\al_T|\al_0)\right],
$$
$$
\Psi_\ep(\al_T|\al_0)=Y_{SO(4,1)}(\al_T)Y^*_{SO(4,1)}(\al_0)
\exp\left\{\imath P\xi\left(\tanh(\xi_T)-\tanh(\xi_0)\right)\right\},
$$
where $A^{(\pm)}$ are the operators of the creation and annihilation of the
Universe, $Y_{SO(4,1)}$ is a unitary irreducible representation of the
$SO(4,1)$
group. This wave function reproduces the physical picture of the Misner
anisotropic Universe (\ref{mis}) discussed above in section 4.

In the large time limit, the $SO(4,1)$ symmetry is broken, the Kazimir
operator term disappears in comparison with the mass term.
 In this case, the masses of elementary particles in the Weinberg-Salam
model are determined by the fixed values of angles of the scale-scalar
field mixing
and the ADM-observer gets the Friedmann cosmological models of radiation
and dust, considered above.

\vspace{0.3cm}

\section{Conclusion}

We have shown that including the time surface term in the canonical
Hamiltonian formulation of GR helps us to extract the time-like variable
and its conjugated momentum from the extended phase space. By taking
into account the time surface term we represented here the new  version
of the Dirac - ADM Hamiltonian formalism for general relativity in the
reduced phase space  with the Schr\"odinger - like equation for a wave
function describing the quantum evolution of the Universe. This evolution
coincides with the Friedmann classical evolution of the dust filled Universe
and shows that in GR like in special relativity there are two distinguished
invariant time variables:
{\bf the "phase time" of the ADM-observer (who constructs the Hamiltonian and
 measures the time as a phase of the wave function of the expanding Universe )
and the geometrical time of the Friedmann observer
(who measures the time as an invariant proper interval and observes
this expansion on the earth)}.

In special relativity, the corresponding times are connected by
the Lorentz transformation and they coincide only in the case
when the rest frame of the Einstein observer coincides with the rest frame
of a particle.
Now the main question is to find  the corresponding transformation
from the rest frame of the ADM - observer to the Friedmann one.

\vspace{0.3cm}
{\bf Acknowledgment}
\vspace{0.3cm}

The authors  thank  Profs. A. Ashtekar,
B M. Barbashov, G.T. Horowits, V.G. Kadyshevsky, K. Kuchar,
 D.A.Kirzhnitz, L.N. Lipatov, D.V.Volkov for useful discussions.
One of the authors (V.N.P.) acknowledges the hospitality of the
International Centre for Theoretical Physics in Trieste where this paper
was finished.

\end{document}